\documentclass[aps,prd,twocolumn,superscriptaddress,nofootinbib]{revtex4}
\pdfoutput=1  
\usepackage{graphicx,color}
\usepackage{appendix}
\usepackage{latexsym,amsmath,amssymb,graphicx,booktabs}
\usepackage{epsfig,latexsym}
\usepackage{hyperref}

\definecolor{MyBlue}{rgb}{0.15,0.15,0.70}

\hypersetup{
colorlinks=true,
citecolor=MyBlue,
linkcolor=MyBlue,
urlcolor=MyBlue
}

\usepackage{amssymb}
\usepackage{amsmath}
\usepackage{amsfonts}
\usepackage{upgreek}
\usepackage{latexsym}

\newcommand{\iBox}{\Box^{-1}}

\renewcommand\({\left(}
\renewcommand\){\right)}
\renewcommand\[{\left[}
\renewcommand\]{\right]}

\newcommand\n{{\mbox {\boldmath $\nabla$}}}
\newcommand{\ra}{\rightarrow}

\def\lsim{\raise 0.4ex\hbox{$<$}\kern -0.8em\lower 0.62
ex\hbox{$\sim$}}

\def\gsim{\raise 0.4ex\hbox{$>$}\kern -0.7em\lower 0.62
ex\hbox{$\sim$}}

\def\lbar{{\hbox{$\lambda$}\kern -0.7em\raise 0.6ex
\hbox{$-$}}}

\newcommand\eq[1]{eq.~(\ref{#1})}
\newcommand\eqs[2]{eqs.~(\ref{#1}) and (\ref{#2})}

\newcommand\eqst[2]{eqs.~(\ref{#1})--(\ref{#2})}
\newcommand\Eqst[2]{Eqs.~(\ref{#1})--(\ref{#2})}
\newcommand\pa{\partial}
\newcommand\p{\partial}

\newcommand\ee{\end{equation}}
\newcommand\be{\begin{equation}}
\def\bea{\begin{array}}
\def\eea{\end{array}}\def\ea{\end{array}}
\newcommand\ees{\end{eqnarray}}
\newcommand\bees{\begin{eqnarray}}
\def\nn{\nonumber}

\def\a{\alpha}
\def\b{\beta}

\def\s{\sigma}
\def\g{\gamma}

\def\d{\delta}

\def\dslash{\hspace{-1mm}\not{\hbox{\kern-2pt $\partial$}}}
\def\Dslash{\not{\hbox{\kern-4pt $D$}}}
\def\pslash{\not{\hbox{\kern-2.1pt $p$}}}
\def\kslash{\not{\hbox{\kern-2.3pt $k$}}}
\def\qslash{\not{\hbox{\kern-2.3pt $q$}}}

\newcommand{\vk}{{\bf k}}
\newcommand{\vx}{{\bf x}}

\def\p1{{\bf p}_1}
\def\p2{{\bf p}_2}
\def\k1{{\bf k}_1}
\def\k2{{\bf k}_2}

\newcommand{\emn}{\eta_{\mu\nu}}

\newcommand{\eMN}{\eta^{\mu\nu}}
\newcommand{\eRS}{\eta^{\rho\sigma}}
\newcommand{\eMR}{\eta^{\mu\rho}}
\newcommand{\eNS}{\eta^{\nu\sigma}}
\newcommand{\eMS}{\eta^{\mu\sigma}}
\newcommand{\eNR}{\eta^{\nu\rho}}

\newcommand{\gmn}{g_{\mu\nu}}

\newcommand{\hmn}{h_{\mu\nu}}
\newcommand{\hrs}{h_{\rho\sigma}}

\newcommand{\pam}{\pa_{\mu}}

\newcommand{\pan}{\pa_{\nu}}
\newcommand{\parho}{\pa_{\rho}}

\newcommand{\paM}{\pa^{\mu}}
\newcommand{\paN}{\pa^{\nu}}
\newcommand{\paR}{\pa^{\rho}}

\newcommand{\Rmn}{R_{\mu\nu}}
\newcommand{\Gmn}{G_{\mu\nu}}
\newcommand{\RMN}{R^{\mu\nu}}

\newcommand{\Tmn}{T_{\mu\nu}}

\newcommand{\TMN}{T^{\mu\nu}}

\newcommand{\dddM}{\kern 0.2em \raise 1.9ex\hbox{$...$}\kern -1.0em \hbox{$M$}}
\newcommand{\dddQ}{\kern 0.2em \raise 1.9ex\hbox{$...$}\kern -1.0em \hbox{$Q$}}
\newcommand{\dddI}{\kern 0.2em \raise 1.9ex\hbox{$...$}\kern -1.0em\hbox{$I$}}
\newcommand{\dddJ}{\kern 0.2em \raise 1.9ex\hbox{$...$}\kern-1.0em
\hbox{$J$}}
\newcommand{\dddcalJ}{\kern 0.2em \raise 1.9ex\hbox{$...$}\kern-1.0em
\hbox{${\cal J}$}}

\newcommand{\dddO}{\kern 0.2em \raise 1.9ex\hbox{$...$}\kern -1.0em
\hbox{${\cal O}$}}
\def\dddz{\raise 1.5ex\hbox{$...$}\kern -0.8em \hbox{$z$}}
\def\dddd{\raise 1.8ex\hbox{$...$}\kern -0.8em \hbox{$d$}}
\def\dddbd{\raise 1.8ex\hbox{$...$}\kern -0.8em \hbox{${\bf d}$}}
\def\ddbd{\raise 1.8ex\hbox{$..$}\kern -0.8em \hbox{${\bf d}$}}
\def\dddx{\raise 1.6ex\hbox{$...$}\kern -0.8em \hbox{$x$}}

\newcommand{\Sch}{Schwarzschild }

\newcommand{\mpl}{M_{\rm Pl}}

\newcommand{\ode}{\Omega_{\rm DE}}
\newcommand{\oma}{\Omega_{M}}
\newcommand{\ora}{\Omega_{R}}

\newcommand{\rde}{\rho_{\rm DE}}

\begin{document}

\title{Nonlocal gravity and dark energy}

\author{Michele Maggiore}  

\affiliation{D\'epartement de Physique Th\'eorique and Center for Astroparticle Physics, 
Universit\'e de Gen\`eve, 24 quai Ansermet, CH--1211 Gen\`eve 4, Switzerland}

\author{Michele Mancarella}
\affiliation{Dipartimento di Fisica``Enrico Fermi", Universit\`a di Pisa, Largo Bruno Pontecorvo, 3, 56127 Pisa, Italy.}

\begin{abstract}

We discuss a nonlocal modification of gravity obtained adding a term $m^2 R\,\Box^{-2}R$ to the Einstein-Hilbert action. We find that the  mass parameter $m$ only affects the non-radiative sector of the theory, while the graviton remains massless, there is no propagating ghost-like degree of freedom, no vDVZ discontinuity, and  no Vainshtein radius below which the theory becomes strongly coupled.
For $m={\cal O}(H_0)$ the theory therefore recovers all successes of GR at solar system and lab scales, and only deviates from it at cosmological scales.  We examine the cosmological consequences of the model and we find that it  automatically  generates a dynamical dark energy and a self-accelerating evolution. After fixing our only free parameter $m$ so to reproduce the observed value of the dark energy density today, we get a pure prediction for the dark energy equation of state,
$w_ {\rm DE}\simeq-1.14$. This value is consistent with the existing data,  and could also resolve the possible tension between the Planck data and local measurements of the Hubble parameter.

\end{abstract}

\maketitle

\section{Introduction}

The experimental observation of the accelerated expansion of the Universe
\cite{Riess:1998cb,Perlmutter:1998np} has stimulated an intense search for modifications of General Relativity (GR) at cosmological scales. The construction of a consistent infrared deformation of GR turns out however to be extremely challenging. A natural way to proceed is to introduce a mass scale $m$ of the order of the present value of the Hubble parameter $H_0$. However such attempts must face a number of difficulties, related to the possible appearance of new ghost-like degrees of freedom (or of ghostlike excitations over non-trivial backgrounds), the appearance of classical or quantum strong coupling regimes, potential problems with causality, while 
at present it is also unclear whether acceptable cosmological solutions emerge
 (see \cite{Hinterbichler:2011tt,deRham:2014zqa} for  recent  reviews). 
 
In a recent series of papers \cite{Jaccard:2013gla,Maggiore:2013mea,Foffa:2013vma,Foffa:2013sma,MMAK} our group has proposed an approach in which a mass parameter enters the theory as the coefficient of a suitable nonlocal term.  
At the level of the general idea,  our approach was inspired by the observation that nonlocal operators provide a way of writing  a mass term, both in massive electrodynamics and in linearized massive gravity, without breaking the gauge invariance of the massless 
theory~\cite{Dvali:2006su} and they can also play an important cosmological role through the degravitation mechanism  \cite{ArkaniHamed:2002fu}. In practice, this general idea can be implemented in different ways. The one closest to the original degravitation idea involves the addition of a term  $m^2(\iBox\Gmn)^{\rm T}$ to the Einstein equations
\cite{Jaccard:2013gla}. The  superscript T  denotes the extraction of the transverse part, which is necessary for consistency with energy-momentum conservation (see also
\cite{Porrati:2002cp} for related ideas). It was then realized in 
\cite{Maggiore:2013mea,Foffa:2013vma,Modesto:2013jea} that such tensor nonlocalities generate instabilities in the cosmological evolution. Similar conclusions were obtained in
\cite{Ferreira:2013tqn} studying a nonlocal model with a term of the form 
$\Rmn\iBox\RMN$ in the action. In refs.~\cite{Maggiore:2013mea,Foffa:2013sma,Foffa:2013vma,MMAK} we have then turned our attention to a model in which a term $m^2(\gmn\iBox R)^ {\rm T} $ is added to the Einstein equations, and we found that it passes a number of tests of theoretical consistency, and  has an interesting cosmological phenomenology. In this paper we turn our attention to a related model, in which again the $\iBox$ operator acts on the Ricci scalar, but which is defined by the nonlocal action
\be\label{S1}
S_{\rm NL}=\frac{1}{16\pi G}\int d^{d+1}x \sqrt{-g}\, 
\[R-\frac{d-1}{4d} m^2R\frac{1}{\Box^2} R\]\, ,
\ee
where  $d$ is the number of spatial dimensions and the factor $(d-1)/4d$ is a convenient normalization of the mass parameter $m$. We will see that this model is quite interesting, both at the theoretical level and for its cosmological consequences.
Non-local cosmological models of different type, not involving a mass scale, have also been  studied recently \cite{Deser:2007jk,Nojiri:2007uq,Jhingan:2008ym,Koivisto:2008xfa,Koivisto:2008dh,Capozziello:2008gu,Elizalde:2011su,Zhang:2011uv,Elizalde:2012ja,Park:2012cp,Bamba:2012ky,Deser:2013uya,Ferreira:2013tqn,Dodelson:2013sma,Woodard:2014iga,Barvinsky:2003kg,Barvinsky:2011hd,Barvinsky:2011rk}.

\section{Equations of motions}

The equations of motion of the theory can be obtained introducing two scalar fields 
\be
U=-\iBox R\, ,
\ee 
and 
\be
S=-\iBox U=\Box^{-2}R\, ,
\ee
and rewriting \eq{S1} as
\bees\label{S2}
S_{\rm NL}&=&(16\pi G)^{-1}\int d^{d+1}x \sqrt{-g}\, \\
&&\times\[ R(1-\mu S)-\xi_1(\Box U+R)-\xi_2 (\Box S+U)\]\nn
\, ,
\ees
where we introduced $\mu=[(d-1)/(4d)]m^2$, and $\xi_1,\xi_2$ are two Lagrange multipliers. The variation is then straightforward and gives (adding also the matter action)
\bees
\Gmn&=&\mu K_{\mu\nu}+8\pi G\Tmn\, ,\label{Gmn}\\
\Box U&=&-R\, ,\qquad
\Box S =-U\, .\label{BoxUS}
\ees
where
\bees
K_{\mu\nu}&=&2S\Gmn-2\n_{\mu}\pan S -2U\gmn+\gmn\parho S\paR U\nn\\
&& -(1/2)\gmn U^2-(\pam S\pan U+\pan S\pam U)\, .
\label{K}
\ees
It is straightforward to check explicitly that $\n^{\mu}K_{\mu\nu}=0$, as it should, since it has been derived from a diff-invariant action.

\section{Radiative and non-radiative degrees of freedom}

A crucial point,
that we already discussed in detail in \cite{Maggiore:2013mea,Foffa:2013sma}
(see also the related discussion in \cite{Koshelev:2008ie,Koivisto:2009jn,Barvinsky:2011rk,Deser:2013uya}) is that, despite the appearance of a Klein-Gordon operator,  \eq{BoxUS} do not describe radiative degrees of freedom. This can be understood as follows. 
In general, an equation such as $\Box U=-R$ is solved by $U=-\iBox R$, where
\be\label{defiBox}
\iBox R= U_{\rm hom}(x)-\int d^{d+1}x'\, \sqrt{-g(x')}\, G(x;x') R(x')\, ,
\ee
with $U_{\rm hom}(x)$ any solution of $\Box U_{\rm hom}=0$, and $G(x;x')$ a Green's function of the $\Box$ operator. The choice of the homogeneous solution is part of the definition of the $\iBox$ operator and therefore of the original nonlocal theory. 
For instance, in a FRW background, on a scalar function $f(t),$  we have $\Box f=-a^{-d}\pa_0(a^d\pa_0f)$. Then one immediately verifies that a possible inversion of the $\iBox$ operator is given by
\be\label{iBoxDW}
(\iBox  R)(t)=-\int_{t_*}^{t} dt'\, \frac{1}{a^d(t')}
\int_{t_*}^{t'}dt''\, a^d(t'') R(t'')\, ,
\ee
where $t_*$ is some initial value of time, that can be taken for instance deep into the radiation dominance (RD) epoch 
(observe that, since in RD the Ricci scalar $R$ vanishes, this definition is independent of the exact value of $t_*$, as long as it is deep in RD). This definition corresponds to a specific choice of $G(x;x') $ and
$U_{\rm hom}(x)$ in \eq{defiBox} and, in particular, with this definition for $t>t_*$ we have $U=0$ if $R=0$, so it corresponds to setting $U_{\rm hom}(t)=0$. In contrast, in the local formulation based on
\eqs{Gmn}{BoxUS}, given a solution for $U$ we can add to it an arbitrary solution of the homogeneous equation $\Box U=0$. However, such a general solution of the local equation is not a solution of the 
integro-differential equation of motion of the original nonlocal model. For instance, with the definition (\ref{iBoxDW}) of the $\iBox$ operator, the original nonlocal model only admits the solution
\be
U(t)=-\int_{t_*}^{t} dt'\, \frac{1}{a^d(t')}
\int_{t_*}^{t'}dt''\, a^d(t'') R(t'')\, ,
\ee
with $U_{\rm hom}(t)=0$. All other solutions of the local formulation are spurious, and do not satisfy the original integro-differential equation.
More generally, whatever definition one takes for
$\iBox$, the corresponding homogeneous solution is uniquely fixed. 
Thus, $U_{\rm hom}(x)$
is not a free field that can be expanded in plane waves which, at the quantum level, would corresponds to creation and annihilation operators of some particle. 

This is a crucial difference between a nonlocal model defined by \eq{S1} and a quintessence model in which the starting point is given by the local equations (\ref{Gmn}) and
(\ref{BoxUS}). If we start from \eqs{Gmn}{BoxUS} and we consider them as the classical  equation of motion of a quantum field theory, the homogeneous equations $\Box U=0$ and $\Box S=0$  have  solutions corresponding to  superposition of plane waves, with arbitrary coefficients $a_{\vk}$ and $a^*_{\vk}$.
Once we move to a quantum description, these would become the creation and annihilation operator of the corresponding particles, so $U$ and $S$ would give rise to radiative degrees of freedom. 
In contrast, if the starting point is the nonlocal theory (\ref{S1}), the auxiliary fields are by definition non-radiative, and there is no quantized field associated to them. 

This is  particularly important because, if it were a radiative degree of freedom,  $U$ would turn out to be a ghost, see \cite{Maggiore:2013mea,Foffa:2013sma}. As discussed in \cite{Foffa:2013sma},
in general a ghost has two quite distinct effects:  at the classical level, it can give rise to classical instability, while at the quantum level it corresponds to a particle with negative energy, which  induces a decay of the vacuum, through processes in which the vacuum decays into ghosts plus normal particles.
Classical instabilities are not necessarily a source of trouble. Actually, 
in  a cosmological context they can even be welcome,  because a phase of accelerated expansion is in a sense  an instability of the classical evolution. Indeed, ghosts have been suggested as models of phantom dark energy~\cite{Caldwell:1999ew,Carroll:2003st}. One must therefore study the classical equations of motion, and see if the consequences of an instability are actually dangerous, or not. This is what we will do in this paper at the level of the background evolution, where we will see that there is indeed a classical instability in the cosmological evolution, which however is nothing but a phase of accelerated expansion, with features well compatible with the observations. In \cite{Dirianetal} we will present the study of the classical cosmological perturbation for this model, and we will see again that the perturbations are well-behaved and consistent with existing data. In contrast, as mentioned above, models in which the $\iBox$ operator is applied to tensors such as $\Gmn$ generates classical instabilities that are inconsistent with an acceptable cosmological evolution \cite{Maggiore:2013mea,Foffa:2013vma,Modesto:2013jea}.

The issue of quantum vacuum decay is different. If, at the quantum level, the vacuum is unstable because of ghost emission, the theory is simply inconsistent. In our case, however, this does not happen because the auxiliary field $U$ is not a radiative field, and does not correspond to particles in the quantum theory.
This advantage over a quintessence model in which one takes the local equations (\ref{Gmn}) and (\ref{BoxUS}) as the starting point, comes however at a price. Namely, a nonlocal theory such as that defined by \eq{S1} cannot be taken as a fundamental theory. It must rather be understood as an effective classical theory, obtained from some fundamental (and local) QFT after a suitable classical or quantum smoothing procedure. As discussed in \cite{Maggiore:2013mea}, an example of situations in which the equation of motion involve an $\iBox$ operator constructed with a retarded Green's function is given by 
the effective equations that 
govern the  dynamics of the in-in matrix elements of quantum fields, such as 
$\langle 0_{\rm in}|\hat{\phi}|0_{\rm in}\rangle$ or $\langle 0_{\rm in}|\hat{g}_{\mu\nu}|0_{\rm in}\rangle$, and encode  quantum corrections  
to the classical dynamics \cite{Jordan:1986ug,Calzetta:1986ey}.
Causal non-local equations  also emerge as a result of a purely classical smoothing, when one separates the dynamics of a system into a long-wavelength and a short-wavelength part,  see e.g. \cite{Carroll:2013oxa} for a recent example in the context of cosmological perturbation theory. 

In this paper we  consider \eq{S1} as an effective nonlocal classical theory, and we  explore its consequences.
It would  of course be of great interest to understand how such a nonlocal model can be derived from a fundamental theory.

\section{Linearization over flat space}

To study the physical content of the nonlocal model and to complement the above discussion, it is useful to  linearize the equations of motion  over Minkowski space. Writing $\gmn=\emn +\hmn$  we get
\be\label{line1}
{\cal E}^{\mu\nu,\rho\sigma}\hrs
-\frac{d-1}{d}\, m^2 P^{\mu\nu}P^{\rho\sigma}
\hrs=-16\pi G\TMN\, ,
\ee
where  ${\cal E}^{\mu\nu,\rho\sigma}$  is the  Lichnerowicz operator  (conventions and definitions are as in \cite{Jaccard:2013gla}),
\be
P^{\mu\nu}=\eMN-\frac{\paM\paN}{\Box}\, ,
\ee
and $\Box$ is now  the   flat-space d'Alembertian. 
This is the same result that was found in \cite{Maggiore:2013mea} linearizing the theory obtained adding directly a term $m^2(\gmn\iBox R)^ {\rm T} $  to the Einstein equations. Thus, the two theories are equivalent at the linearized level. At the fully non-linear level they are however different, as can be seen by comparing the respective equations of motion.\footnote{Observe that in ref.~\cite{Modesto:2013jea} it was studied an action proportional to $\Gmn\Box^{-2}\RMN$ and it was claimed that, below the Planck scale,  the theory reduces to that obtained by adding a term $(\iBox\Gmn)^{\rm T}$  directly to the Einstein equations. By the same token one would conclude that at sub-planckian curvatures the theory 
(\ref{S1}) reduces to obtained adding directly a term $m^2(\gmn\iBox R)^ {\rm T} $  to the Einstein equations. Unfortunately, in both cases the argument is incorrect. The non-linearities (e.g. the term $U^2$ in \eq{K})
are suppressed, with respect to the linear terms,  by a factor ${\cal O}(\iBox R)$, which is just $O(h)$, and not by
${\cal O}(R/\mpl^2)$. They are on the same footing as the usual non-linearities of GR, and contribute whenever we are not close to flat Minkowski space. We will indeed see that the model (\ref{S1}) has cosmological predictions that are numerically different from that of the model obtained adding directly a term $m^2(\gmn\iBox R)^ {\rm T} $  to the Einstein equations,  studied in \cite{Maggiore:2013mea}.}

In order to study what radiative and non-radiative degrees of freedom are described by 
\eq{line1} we proceed as in GR. We  henceforth restrict to $d=3$, we consider first the scalar sector, and we use the diff-invariance of the nonlocal theory to fix the Newtonian gauge 
\be
h_{00}=2\Psi\, ,\qquad
h_{0i}=0\, ,\qquad h_{ij}=2\Phi\d_{ij}\, . 
\ee 
We also write the energy-momentum tensor in the scalar sector as 
\bees
T_{00}&=&\rho\, , \qquad T_{0i}=\pa_i\Sigma\, ,\\
T_{ij}&=&P\d_{ij}+[\pa_i\pa_j-(1/3)\d_{ij}\n^2]\sigma\, . 
\ees
A straightforward generalization of the standard computation performed in GR (see e.g. \cite{Jaccard:2012ut}) 
gives  four independent equations for the four scalar variables $\Phi,\Psi$, $U$ and $S$,
\bees
\n^2\[\Phi-(m^2/6) S\]&=&-4\pi G\rho\, ,\label{dof1}\\
\Phi-\Psi-(m^2/3) S&=& -8\pi G\sigma\, ,\label{dof2}\\
(\Box+m^2)U&=&-8\pi G (\rho-3P)\, ,\label{dof3}\\
\Box S&=&-U\, .\label{dof4}
\ees
Eqs.~(\ref{dof1}) and (\ref{dof2}) show that $\Phi$ and $\Psi$ remain non-radiative, just as in GR. This should be contrasted with what happens when one linearizes massive gravity with a Fierz-Pauli mass term, in which case $\Phi$ becomes a radiative field that satisfies 
$(\Box-m^2)\Phi=0$ \cite{Deser:1966zzb,Alberte:2010it,Jaccard:2012ut}. Furthermore, in local massive gravity with a mass term that does not satisfies  the Fierz-Pauli tuning, in the Lagrangian also  appears a term $(\Box\Phi)^2$ \cite{Jaccard:2012ut}, signaling the presence of a dynamical ghost. In our nonlocal model, in contrast, $\Phi$ and $\Psi$ satisfy Poisson equations and therefore remain non-radiative.
The equations for $U$ and $S$ might fool us to believe that we have two radiative scalars. However, 
\eq{dof3} is just the linearization of $\Box U=-R$ where, as we have discussed, the radiative solution is a spurious one, introduced when the original nonlocal model is written  in a local form using the auxiliary fields $U$ and $S$. In a quantum treatment, there are no annihilation and creation operators associated to them, and they do not represent radiative degrees of freedom of the original nonlocal theory (see also the discussion in \cite{Foffa:2013sma}). 
Observe that the argument on the absence of radiative ghost-like degrees of freedom is not restricted to the linearized approximation. The full non-linear equations (\ref{BoxUS}) by definition must be supplemented with a given fixed choice of the homogeneous solutions, so they never describe  propagating fields.

The full content of the theory beyond the scalar sector can be obtained from  the computation of ref.~\cite{Maggiore:2013mea}
of the matter-matter interaction mediated by the theory (\ref{line1}). In $d=3$ the result  is proportional to 
\bees
&& \tilde{T}_{\mu\nu}(-k)\frac{1}{2k^2}\, 
\( \eMR\eNS +\eMS\eNR-\eMN\eRS \)\tilde{T}_{\rho\sigma}(k) \nn\\
&&+\frac{1}{6}\tilde{T}(-k)\( \frac{1}{k^2}-\frac{1}{k^2-m^2}\)\tilde{T}(k)\, .
\label{TDT}
\ees
The term in the first line is the usual GR result  due to the exchange of a massless graviton. The term in the second line is due to the fields $U$ and $S$. If $U$ were a radiative field, its contribution would correspond to that of a ghost, and at the quantum level the vacuum would get destabilized. However, the previous analysis show that there is no radiative degree of freedom associated to these terms.

\section{Absence of vDVZ discontinuity and of a Vainsthein mechanism}

Eq.~(\ref{TDT}) shows that, in the limit $m\ra 0$, the matter-matter interaction reduces smoothly to that of GR. Therefore there is no vDVZ discontinuity,
and no Vainshtein mechanism is needed. Of course, by itself this  does not necessarily mean that non-linearities will remain small down to the \Sch radius $r_S$, where also the classical non-linearities of GR get large. However, this can be checked computing the metric generated by static sources in the nonlocal theory. This computation has been performed in detail in
\cite{MMAK} for the model defined adding  a term $m^2(\gmn\iBox R)^ {\rm T} $  to the Einstein equations, and can be simply adapted to our case. We write the most general static spherically symmetric metric in the form
\be\label{ds2}
ds^2=-e^{2\alpha(r)} dt^2+e^{2\beta(r)}dr^2 +r^2(d\theta^2+\sin^2\theta\,  d\phi^2)\, .
\ee
In GR two independent equations for $\a$ and $\b$ are usually obtained  taking the combinations 
$e^{2(\b-\a)}R_{00}+R_{11}$ and $R_{22}$ (see e.g. \cite{Carroll:1997ar}). 
In our nonlocal theory, using \eq{Gmn} we get, respectively 
\be\label{eqab1}
(1-2\mu S)(\a'+\b')=-\mu r [S''-(\a'+\b'-U')S']\, ,
\ee
(where $f'\equiv df/dr$), and
\bees
&&\hspace*{-10mm}(1-2\mu S)\left\{1+e^{-2\b}[r(\b'-\a')-1]\right\}\nn\\
&&\hspace*{-5mm}=\mu r^2(U+U^2/2)-2\mu r e^{-2\b}S'\, ,\label{eqab2}
\ees
which reduce to their GR counterparts for $\mu=0$. Finally, in the metric (\ref{ds2}) \eq{BoxUS} becomes
\bees
&&r^2U''+[2r+(\a'-\b')r^2]U'\label{eqU}\\
&&=-2 e^{2\b}+2[1+2r(\a'-\b')+r^2(\a''+{\a'}^2-\a'\b')]\, ,\nn\\
&&S'' +(\a'-\b' +2/r)S'=-e^{2\b}U\, .\label{eqS}
\ees
\Eqst{eqab1}{eqS} provide four independent equations for the four functions $\a,\b,U,S$. As discussed in \cite{MMAK}, we can study these equations with two different expansions: in the region $r\ll m^{-1}$ we can perform a low-$m$ expansion, in which
we solve the equation iteratively taking $m$ as a small expansion parameter. The solution in the region $r\gg r_S$, with no limitation of the parameter $mr$, can instead be obtained 
considering the effect of the source  as a perturbation of  Minkowski space, adapting the standard analysis performed in GR to recover the Newtonian limit. The low-$m$ expansion is valid for $mr\ll 1$ while the Newtonian analysis is valid for $r\gg r_S$. The two expansions therefore have an overlapping domain of validity
$r_S\ll r\ll m^{-1}$, where they can be matched, and this allows us to fix uniquely all the coefficients that appears in the solutions, see the discussion in \cite{MMAK}.

Repeating for our model the computations performed in \cite{MMAK}
we find that, to first order in the low-$m$ expansion and for $r\gg r_S$, the result for $\a$ and $\b$ is the same as that found in \cite{MMAK}. Concerning the Newtonian expansion, we have already seen that the theory (\ref{S1}) and that based on the
$m^2(\gmn\iBox R)^ {\rm T}$ term become identical when they are both linearized over Minkowski space. Thus, also the result in this limit is the same, and the matching procedure discussed in \cite{MMAK} goes through without any modification. Thus, we can simply read the result from \cite{MMAK}: writing $A(r)=e^{2\a}$ and
$B(r)=e^{2\b}$, the solution for $r\gg r_S$ (and $mr$ generic) is
\bees
\hspace*{-2mm}A(r)&=&1-\frac{r_S}{r}\[1+\frac{1}{3}(1-\cos mr)\]\, ,\label{NewtA}\\
\hspace*{-2mm}B(r)&=&1+\frac{r_S}{r}\[1-\frac{1}{3}(1-\cos mr-mr \sin mr)
\]\hspace{-1mm}.\label{NewtB}
\ees
In particular, for $r_S\ll r\ll m^{-1}$ we have
\be\label{Afinal}
A(r)\simeq1-\frac{r_S}{r}\(1+\frac{m^2 r^2}{6}\)\, ,
\ee
and $B(r)\simeq 1/A(r)$. This should be compared with the analogous result
obtained in
massive gravity, when one considers the Einstein-Hilbert action plus a Fierz-Pauli mass term, which reads \cite{Vainshtein:1972sx,Hinterbichler:2011tt}
\be\label{AFP}
A(r)=1-\frac{4}{3}\frac{r_S}{r}\(1-\frac{r_S}{12m^4r^5}\)\, .
\ee
The factor $4/3$ in front of $r_S/r$  gives rise to the vDVZ discontinuity. In contrast, no vDVZ discontinuity is present in
\eq{Afinal}. Furthermore, in \eq{AFP} the linearized expansions breaks down
for 
$r$ below the 
Vainshtein radius  $r_V=(GM/m^4)^{1/5}$, while in \eq{Afinal}  the correction becomes smaller and smaller as $r$ decreases.  Thus   the theory
(\ref{S1}) (as well as   the theory defined adding a term $m^2(\gmn\iBox R)^ {\rm T} $ to the Einstein equations) remain linear down to $r\sim r_S$, where eventually also GR becomes non-linear. This means that,
taking $m\sim H_0$, these nonlocal theories pass with flying colors all solar system tests. We have found that, for $r\ll m^{-1}$, the corrections to the GR result are
$1+{\cal O}(m^2r^2)$. For 
$m\sim H_0$ and $r\sim 1\, {\rm a.u.}$  we have
$m^2r^2\sim10^{-30}$, and the predictions of these nonlocal theories are indistinguishable from that of GR.

\section{Cosmological evolution and dark energy} 

We next study the cosmological consequences of the model, at the level of background evolution (the corresponding study of cosmological perturbations will be presented in
\cite{Dirianetal}).
We consider a flat FRW metric  
\be
ds^2=-dt^2+a^2(t)d\vx^2\, ,
\ee 
in  $d=3$. We introduce $W(t)=H^2(t)S(t)$ and $h(t)=H(t)/H_0$, where  $H(t)=\dot{a}/a$ and $H_0$ is the present value of the Hubble parameter. We use $x=\ln a$ to parametrize the temporal evolution, and henceforth $f'\equiv df/dx$. From \eqs{Gmn}{BoxUS} we get
\bees
&&h^2(x)=\Omega_M e^{-3x}+\Omega_R e^{-4x}+\g Y\label{syh}\\
&&U''+(3+\zeta) U'=6(2+\zeta)\, ,\label{syU}\\
&&W''+3(1-\zeta) W'-2(\zeta'+3\zeta-\zeta^2)W= U\, ,\label{syW}
\ees
where $\gamma= m^2/(9H_0^2)$, $\zeta=h'/h$ and
\be
Y\equiv \frac{1}{2}W'(6-U') +W (3-6\zeta+\zeta U')+\frac{1}{4}U^2\, .
\ee
We see that there is an effective dark energy density $\rde=\rho_0\gamma Y$.
As in \cite{Maggiore:2013mea}, we 
can first study the equations perturbatively, assuming that in the early Universe the contribution of $U,V$ to $\zeta$ is negligible, 
and we then check a posteriori the self-consistency of the procedure. In this case, in each given era $\zeta(x)$ can be approximated by a constant $\zeta_0$, with 
$\zeta_0=\{-2,-3/2,0\}$ in RD, MD and in a De~Sitter inflationary epoch, respectively.
We find that the inhomogeneous solutions for $U$ and $W$  are both linear in $x$, and therefore as $x\ra-\infty$ their contribution is indeed negligible with respect to the terms $\Omega_M e^{-3x}$ and $\Omega_R e^{-4x}$. Furthermore, the corresponding homogeneous solution for $U$ is $u_0
+u_1 e^{-(3+\zeta_0)x}$, while for $W$  is
$w_1e^{-(3-\zeta_0)x}+w_2 e^{2\zeta_0x}$.
In the early Universe we have $-2\leq \zeta_0\leq 0$ and all these terms are either constant or exponentially decreasing, which means that the 
solutions for both $U$ and $W$ are stable  in MD, RD, as well as in a previous  inflationary stage. In contrast, the homogeneous solutions of the model constructed with $(\gmn\iBox R)^T$ are stable in MD and RD, but not in an inflationary stage
\cite{Maggiore:2013mea,Foffa:2013vma}.
 
 \begin{figure}[t]
\centering
\includegraphics[width=0.48\columnwidth]{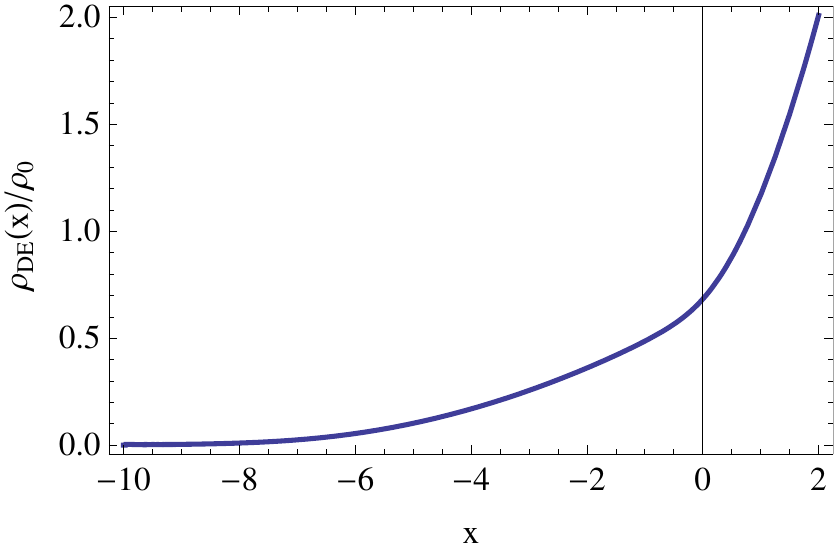}
\includegraphics[width=0.48\columnwidth]{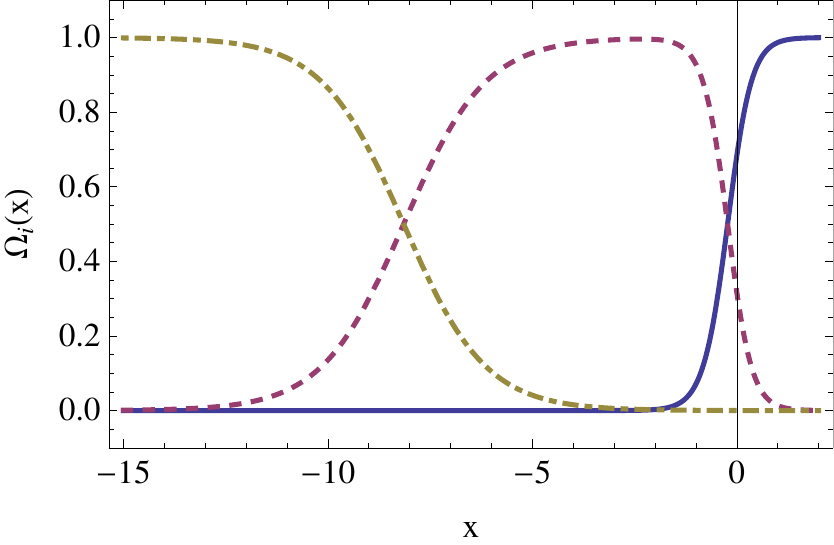}
\caption{\label{fig:rhoDE} Left panel: the function $\g Y(x)=\rde(x)/\rho_0$, against $x=\ln a$.
Right panel: the quantities 
$\Omega_R(x)$ (brown, dot-dashed), $\Omega_M(x)$ (red, dashed)
and $\Omega_{\rm DE}(x)$ (blue, solid line).
}
\end{figure}
\begin{figure}[t]
\centering
\includegraphics[width=0.48\columnwidth]{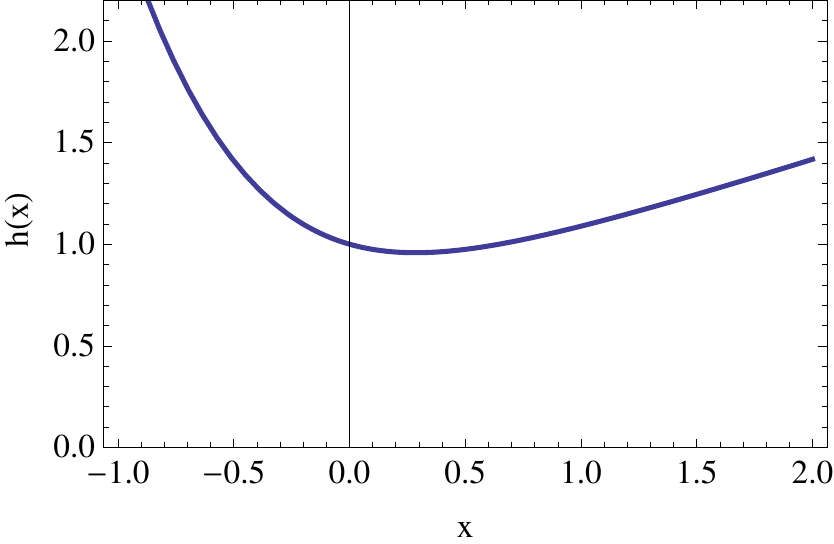}
\includegraphics[width=0.48\columnwidth]{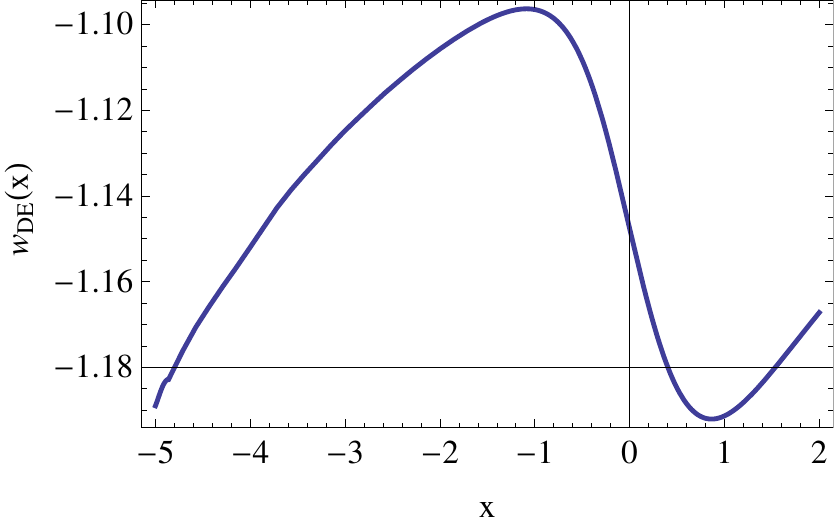}
\caption{\label{fig:w} 
Left panel:  the normalized Hubble parameter $h(x)=H(x)/H_0$.
Right panel: the EOS parameter $w_{\rm DE}(x)$.
}
\end{figure}

\begin{figure}[th]
\centering
\includegraphics[width=0.5\columnwidth]{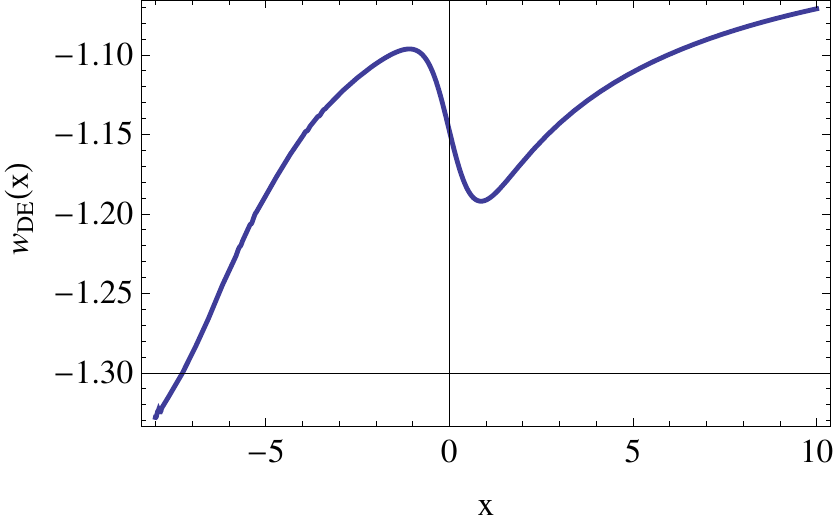}
\caption{\label{fig:wlarge} 
The EOS parameter $w_{\rm DE}(x)$ on a larger horizontal scale.
}
\end{figure}

The result of the numerical integration of \eqst{syh}{syW} is shown in 
Figs.~\ref{fig:rhoDE}-\ref{fig:wlarge}. We start the integration in RD (matter-radiation equilibrium is a $x\simeq -8.1$) with initial conditions $U=S=0$, as in
\cite{Maggiore:2013mea}. On the left panel of Fig.~1 we show  the effective dark energy density
$\rde=\rho_0\gamma Y$. We see that it starts from zero in RD and then grows during MD. Choosing $\gamma\simeq 0.00891$ (corresponding to $m\simeq 0.283 H_0$) we reproduce the observed value $\ode\simeq 0.68$ today. The quantities 
$\Omega_R(x)=\rho_R(x)/\rho_{\rm tot}(x)$, $\Omega_M(x)=\rho_M(x)/\rho_{\rm tot}(x)$
and $\Omega_{\rm DE}(x)=\rho_{\rm DE}(x)/\rho_{\rm tot}(x)$ are shown on the  
right panel of 
Fig.~1.  

In  Fig.~\ref{fig:w} (left panel) we 
see that $h(x)$ becomes a growing function when the DE density begins to dominate. Having fixed $\g$ so that $\ode\simeq 0.68$ we have  no more free parameters, and we then obtain a pure prediction for the dark energy equation of state parameter $w_{\rm DE}$, defined from 
\be\label{consrho}
{\rho}'_{\rm DE}+3(1+w_{\rm DE})\rho_{\rm DE}=0\, .
\ee
The result is shown on the right panel in Fig.~\ref{fig:w} and, on a larger horizontal scale,
in Fig.~\ref{fig:wlarge}, which shows that in the  asymptotic future $w_{\rm DE}\ra -1$.
For comparison with the observations the most relevant region is  the recent past, where the DE density start to become important. Comparing with the standard fit of the form \cite{Chevallier:2000qy,Linder:2002et}
\be
w_{\rm DE}(a)= w_0+(1-a) w_a\, ,
\ee
(where $a(x)=e^x$) in the region $-1<x<0$, we find 
the best-fit values 
\be\label{predw0wa}
w_0 = -1.144\, ,  \qquad w_a = 0.084\, .
\ee
The fact that the EOS turns out to be on the phantom side is a general property of these nonlocal models, due to the fact the DE density starts from zero in RD and then grows during MD. Thus, in this regime  $\rde >0$ and ${\rho}'_{\rm DE}>0$, and  then \eq{consrho} gives $(1+w_{\rm DE})<0$. 
The exact values in \eq{predw0wa} depend of course on the value chosen for the matter density today, $\oma$, or equivalently on the value $\ode =1-\oma-\ora$ that we require at the present time. In turn, the value of $\oma$ predicted by this nonlocal model should be determined self-consistently from a global fit to the data, which takes into account the specific form of the perturbations in this model.
However, varying $\oma$ within the rather broad range $\Omega_M\in [0.20,0.36]$ we find that
$w_0$ remains within the relatively narrow interval $[-1.165,-1.135]$, while
$w_a\in [0.07,0.11]$, so the prediction for these quantities is quite stable (see \cite{Dirianetal} for details).

The numerical values in  (\ref{predw0wa})
are quite interesting, considering that the   result of Planck+WP+SNLS
 for a constant $w_{\rm DE}$ (which is appropriate to our case since we predict $|w_a|\ll 1$)  is  
 \be
 w_{\rm DE}=-1.13^{+0.13}_{-0.14}\, ,
 \ee
at 95\% c.l. \cite{Ade:2013zuv}. Observe also that the Pan-STARRS1 data, combined with 
BAO+Planck+$H_0$, give~\cite{Rest:2013bya} 
\be
w_{\rm DE}=-1.186^{+0.076}_{-0.065}\, ,
\ee 
while, when combined with 
WMAP9 instead of Planck,  give~\cite{Rest:2013bya} 
\be
w_{\rm DE}=-1.142^{+0.076}_{-0.087}\, .
\ee  
As discussed in detail in \cite{Shafer:2013pxa}, the result depends on the prior on $H_0$, and for a prior $H_0\gsim 71$ km s/Mpc, 
at the  $2\s$ level one can state that either the SNLS and  Pan-STARRS1 data both
have systematics that remain unaccounted for, or  the DE equation of state is indeed phantom. Of course, it should be kept in mind that the above experimental values have been inferred from the data  assuming wCDM 
as a cosmological model (which assumes no dark energy perturbations), and again a precise comparison of our model with the data requires the inclusion of the specific form of the perturbations of the nonlocal model. 

Another elements that makes the values (\ref{predw0wa}) potentially interesting is that,
as discussed in the official Planck analysis \cite{Ade:2013zuv}, in the framework of 
$\Lambda$CDM there is a tension between 
 the value of $H_0$  derived from the Planck measurement  and that derived from  direct measurements in the local Universe \cite{Riess:2011yx,Freedman:2012ny}. It has been argued that the discrepancy could be resolved at the level of data analysis \cite{Efstathiou:2013via}. It is however in principle possible that it could  rather be a signal of deviations from $\Lambda$CDM.  Ref.~\cite{Verde:2013wza} has studied the impact of  various extensions of $\Lambda$CDM (such as curvature, neutrino masses, effective neutrino species or $w_{\rm DE}$) on such a discrepancy. It has been found that the only parameter that can reduce the tension to a statistically non-significant value is indeed $w_{\rm DE}$, and this requires a value of $w_{\rm DE}$ approximately in the range
$-1.3< w_{\rm DE}<-1.1$. Our prediction (\ref{predw0wa}) is
therefore  able to bring this discrepancy down to a statistically not significant value.
It is quite remarkable that such a value of  $w_{\rm DE}$ is predicted by a relatively simple and theoretically consistent modification of GR.

\vspace{5mm}
\noindent
{\bf Acknowledgements}.
We thank  Yves Dirian, Stefano Foffa, Maud Jaccard, Alex Kehagias, Ermis Mitsou and Licia Verde for  useful discussions. The work of Michele Maggiore is  supported by the Fonds National Suisse.

\bibliography{myrefs_massive}

\begin{thebibliography}{53}
\expandafter\ifx\csname natexlab\endcsname\relax\def\natexlab#1{#1}\fi
\expandafter\ifx\csname bibnamefont\endcsname\relax
  \def\bibnamefont#1{#1}\fi
\expandafter\ifx\csname bibfnamefont\endcsname\relax
  \def\bibfnamefont#1{#1}\fi
\expandafter\ifx\csname citenamefont\endcsname\relax
  \def\citenamefont#1{#1}\fi
\expandafter\ifx\csname url\endcsname\relax
  \def\url#1{\texttt{#1}}\fi
\expandafter\ifx\csname urlprefix\endcsname\relax\def\urlprefix{URL }\fi
\providecommand{\bibinfo}[2]{#2}
\providecommand{\eprint}[2][]{\url{#2}}

\bibitem[{\citenamefont{Riess et~al.}(1998)}]{Riess:1998cb}
\bibinfo{author}{\bibfnamefont{A.~G.} \bibnamefont{Riess}} \bibnamefont{et~al.}
  (\bibinfo{collaboration}{Supernova Search Team}),
  \bibinfo{journal}{Astron.J.} \textbf{\bibinfo{volume}{116}},
  \bibinfo{pages}{1009} (\bibinfo{year}{1998}), \eprint{astro-ph/9805201}.

\bibitem[{\citenamefont{Perlmutter et~al.}(1999)}]{Perlmutter:1998np}
\bibinfo{author}{\bibfnamefont{S.}~\bibnamefont{Perlmutter}}
  \bibnamefont{et~al.} (\bibinfo{collaboration}{Supernova Cosmology Project}),
  \bibinfo{journal}{Astrophys.J.} \textbf{\bibinfo{volume}{517}},
  \bibinfo{pages}{565} (\bibinfo{year}{1999}), \eprint{astro-ph/9812133}.

\bibitem[{\citenamefont{Hinterbichler}(2012)}]{Hinterbichler:2011tt}
\bibinfo{author}{\bibfnamefont{K.}~\bibnamefont{Hinterbichler}},
  \bibinfo{journal}{Rev.Mod.Phys.} \textbf{\bibinfo{volume}{84}},
  \bibinfo{pages}{671} (\bibinfo{year}{2012}), \eprint{1105.3735}.

\bibitem[{\citenamefont{de~Rham}(2014)}]{deRham:2014zqa}
\bibinfo{author}{\bibfnamefont{C.}~\bibnamefont{de~Rham}}
  (\bibinfo{year}{2014}), \eprint{1401.4173}.

\bibitem[{\citenamefont{Jaccard
  et~al.}(2013{\natexlab{a}})\citenamefont{Jaccard, Maggiore, and
  Mitsou}}]{Jaccard:2013gla}
\bibinfo{author}{\bibfnamefont{M.}~\bibnamefont{Jaccard}},
  \bibinfo{author}{\bibfnamefont{M.}~\bibnamefont{Maggiore}}, \bibnamefont{and}
  \bibinfo{author}{\bibfnamefont{E.}~\bibnamefont{Mitsou}},
  \bibinfo{journal}{Phys.Rev.} \textbf{\bibinfo{volume}{D88}},
  \bibinfo{pages}{044033} (\bibinfo{year}{2013}{\natexlab{a}}),
  \eprint{1305.3034}.

\bibitem[{\citenamefont{Maggiore}(2014)}]{Maggiore:2013mea}
\bibinfo{author}{\bibfnamefont{M.}~\bibnamefont{Maggiore}},
  \bibinfo{journal}{Phys.Rev.} \textbf{\bibinfo{volume}{D89}},
  \bibinfo{pages}{043008} (\bibinfo{year}{2014}), \eprint{1307.3898}.

\bibitem[{\citenamefont{Foffa et~al.}(2013{\natexlab{a}})\citenamefont{Foffa,
  Maggiore, and Mitsou}}]{Foffa:2013vma}
\bibinfo{author}{\bibfnamefont{S.}~\bibnamefont{Foffa}},
  \bibinfo{author}{\bibfnamefont{M.}~\bibnamefont{Maggiore}}, \bibnamefont{and}
  \bibinfo{author}{\bibfnamefont{E.}~\bibnamefont{Mitsou}}
  (\bibinfo{year}{2013}{\natexlab{a}}), \eprint{1311.3435}.

\bibitem[{\citenamefont{Foffa et~al.}(2013{\natexlab{b}})\citenamefont{Foffa,
  Maggiore, and Mitsou}}]{Foffa:2013sma}
\bibinfo{author}{\bibfnamefont{S.}~\bibnamefont{Foffa}},
  \bibinfo{author}{\bibfnamefont{M.}~\bibnamefont{Maggiore}}, \bibnamefont{and}
  \bibinfo{author}{\bibfnamefont{E.}~\bibnamefont{Mitsou}}
  (\bibinfo{year}{2013}{\natexlab{b}}), \eprint{1311.3421}.

\bibitem[{\citenamefont{Kehagias and Maggiore}(2014)}]{MMAK}
\bibinfo{author}{\bibfnamefont{A.}~\bibnamefont{Kehagias}} \bibnamefont{and}
  \bibinfo{author}{\bibfnamefont{M.}~\bibnamefont{Maggiore}}
  (\bibinfo{year}{2014}), \eprint{1401.8289}.

\bibitem[{\citenamefont{Dvali}(2006)}]{Dvali:2006su}
\bibinfo{author}{\bibfnamefont{G.}~\bibnamefont{Dvali}}, \bibinfo{journal}{New
  J.Phys.} \textbf{\bibinfo{volume}{8}}, \bibinfo{pages}{326}
  (\bibinfo{year}{2006}), \eprint{hep-th/0610013}.

\bibitem[{\citenamefont{Arkani-Hamed et~al.}(2002)\citenamefont{Arkani-Hamed,
  Dimopoulos, Dvali, and Gabadadze}}]{ArkaniHamed:2002fu}
\bibinfo{author}{\bibfnamefont{N.}~\bibnamefont{Arkani-Hamed}},
  \bibinfo{author}{\bibfnamefont{S.}~\bibnamefont{Dimopoulos}},
  \bibinfo{author}{\bibfnamefont{G.}~\bibnamefont{Dvali}}, \bibnamefont{and}
  \bibinfo{author}{\bibfnamefont{G.}~\bibnamefont{Gabadadze}}
  (\bibinfo{year}{2002}), \eprint{hep-th/0209227}.

\bibitem[{\citenamefont{Porrati}(2002)}]{Porrati:2002cp}
\bibinfo{author}{\bibfnamefont{M.}~\bibnamefont{Porrati}},
  \bibinfo{journal}{Phys.Lett.} \textbf{\bibinfo{volume}{B534}},
  \bibinfo{pages}{209} (\bibinfo{year}{2002}), \eprint{hep-th/0203014}.

\bibitem[{\citenamefont{Modesto and Tsujikawa}(2013)}]{Modesto:2013jea}
\bibinfo{author}{\bibfnamefont{L.}~\bibnamefont{Modesto}} \bibnamefont{and}
  \bibinfo{author}{\bibfnamefont{S.}~\bibnamefont{Tsujikawa}}
  (\bibinfo{year}{2013}), \eprint{1307.6968}.

\bibitem[{\citenamefont{Ferreira and Maroto}(2013)}]{Ferreira:2013tqn}
\bibinfo{author}{\bibfnamefont{P.~G.} \bibnamefont{Ferreira}} \bibnamefont{and}
  \bibinfo{author}{\bibfnamefont{A.~L.} \bibnamefont{Maroto}}
  (\bibinfo{year}{2013}), \eprint{1310.1238}.

\bibitem[{\citenamefont{Deser and Woodard}(2007)}]{Deser:2007jk}
\bibinfo{author}{\bibfnamefont{S.}~\bibnamefont{Deser}} \bibnamefont{and}
  \bibinfo{author}{\bibfnamefont{R.}~\bibnamefont{Woodard}},
  \bibinfo{journal}{Phys.Rev.Lett.} \textbf{\bibinfo{volume}{99}},
  \bibinfo{pages}{111301} (\bibinfo{year}{2007}), \eprint{0706.2151}.

\bibitem[{\citenamefont{Nojiri and Odintsov}(2008)}]{Nojiri:2007uq}
\bibinfo{author}{\bibfnamefont{S.}~\bibnamefont{Nojiri}} \bibnamefont{and}
  \bibinfo{author}{\bibfnamefont{S.~D.} \bibnamefont{Odintsov}},
  \bibinfo{journal}{Phys.Lett.} \textbf{\bibinfo{volume}{B659}},
  \bibinfo{pages}{821} (\bibinfo{year}{2008}), \eprint{0708.0924}.

\bibitem[{\citenamefont{Jhingan et~al.}(2008)\citenamefont{Jhingan, Nojiri,
  Odintsov, Sami, Thongkool et~al.}}]{Jhingan:2008ym}
\bibinfo{author}{\bibfnamefont{S.}~\bibnamefont{Jhingan}},
  \bibinfo{author}{\bibfnamefont{S.}~\bibnamefont{Nojiri}},
  \bibinfo{author}{\bibfnamefont{S.}~\bibnamefont{Odintsov}},
  \bibinfo{author}{\bibfnamefont{M.}~\bibnamefont{Sami}},
  \bibinfo{author}{\bibfnamefont{I.}~\bibnamefont{Thongkool}},
  \bibnamefont{et~al.}, \bibinfo{journal}{Phys.Lett.}
  \textbf{\bibinfo{volume}{B663}}, \bibinfo{pages}{424} (\bibinfo{year}{2008}),
  \eprint{0803.2613}.

\bibitem[{\citenamefont{Koivisto}(2008{\natexlab{a}})}]{Koivisto:2008xfa}
\bibinfo{author}{\bibfnamefont{T.}~\bibnamefont{Koivisto}},
  \bibinfo{journal}{Phys.Rev.} \textbf{\bibinfo{volume}{D77}},
  \bibinfo{pages}{123513} (\bibinfo{year}{2008}{\natexlab{a}}),
  \eprint{0803.3399}.

\bibitem[{\citenamefont{Koivisto}(2008{\natexlab{b}})}]{Koivisto:2008dh}
\bibinfo{author}{\bibfnamefont{T.}~\bibnamefont{Koivisto}},
  \bibinfo{journal}{Phys.Rev.} \textbf{\bibinfo{volume}{D78}},
  \bibinfo{pages}{123505} (\bibinfo{year}{2008}{\natexlab{b}}),
  \eprint{0807.3778}.

\bibitem[{\citenamefont{Capozziello et~al.}(2009)\citenamefont{Capozziello,
  Elizalde, Nojiri, and Odintsov}}]{Capozziello:2008gu}
\bibinfo{author}{\bibfnamefont{S.}~\bibnamefont{Capozziello}},
  \bibinfo{author}{\bibfnamefont{E.}~\bibnamefont{Elizalde}},
  \bibinfo{author}{\bibfnamefont{S.}~\bibnamefont{Nojiri}}, \bibnamefont{and}
  \bibinfo{author}{\bibfnamefont{S.~D.} \bibnamefont{Odintsov}},
  \bibinfo{journal}{Phys.Lett.} \textbf{\bibinfo{volume}{B671}},
  \bibinfo{pages}{193} (\bibinfo{year}{2009}), \eprint{0809.1535}.

\bibitem[{\citenamefont{Elizalde et~al.}(2012)\citenamefont{Elizalde, Pozdeeva,
  and Vernov}}]{Elizalde:2011su}
\bibinfo{author}{\bibfnamefont{E.}~\bibnamefont{Elizalde}},
  \bibinfo{author}{\bibfnamefont{E.}~\bibnamefont{Pozdeeva}}, \bibnamefont{and}
  \bibinfo{author}{\bibfnamefont{S.~Y.} \bibnamefont{Vernov}},
  \bibinfo{journal}{Phys.Rev.} \textbf{\bibinfo{volume}{D85}},
  \bibinfo{pages}{044002} (\bibinfo{year}{2012}), \eprint{1110.5806}.

\bibitem[{\citenamefont{Zhang and Sasaki}(2012)}]{Zhang:2011uv}
\bibinfo{author}{\bibfnamefont{Y.}~\bibnamefont{Zhang}} \bibnamefont{and}
  \bibinfo{author}{\bibfnamefont{M.}~\bibnamefont{Sasaki}},
  \bibinfo{journal}{Int.J.Mod.Phys.} \textbf{\bibinfo{volume}{D21}},
  \bibinfo{pages}{1250006} (\bibinfo{year}{2012}), \eprint{1108.2112}.

\bibitem[{\citenamefont{Elizalde et~al.}(2013)\citenamefont{Elizalde, Pozdeeva,
  and Vernov}}]{Elizalde:2012ja}
\bibinfo{author}{\bibfnamefont{E.}~\bibnamefont{Elizalde}},
  \bibinfo{author}{\bibfnamefont{E.}~\bibnamefont{Pozdeeva}}, \bibnamefont{and}
  \bibinfo{author}{\bibfnamefont{S.~Y.} \bibnamefont{Vernov}},
  \bibinfo{journal}{Class.Quant.Grav.} \textbf{\bibinfo{volume}{30}},
  \bibinfo{pages}{035002} (\bibinfo{year}{2013}), \eprint{1209.5957}.

\bibitem[{\citenamefont{Park and Dodelson}(2013)}]{Park:2012cp}
\bibinfo{author}{\bibfnamefont{S.}~\bibnamefont{Park}} \bibnamefont{and}
  \bibinfo{author}{\bibfnamefont{S.}~\bibnamefont{Dodelson}},
  \bibinfo{journal}{Phys.Rev.} \textbf{\bibinfo{volume}{D87}},
  \bibinfo{pages}{024003} (\bibinfo{year}{2013}), \eprint{1209.0836}.

\bibitem[{\citenamefont{Bamba et~al.}(2012)\citenamefont{Bamba, Nojiri,
  Odintsov, and Sasaki}}]{Bamba:2012ky}
\bibinfo{author}{\bibfnamefont{K.}~\bibnamefont{Bamba}},
  \bibinfo{author}{\bibfnamefont{S.}~\bibnamefont{Nojiri}},
  \bibinfo{author}{\bibfnamefont{S.~D.} \bibnamefont{Odintsov}},
  \bibnamefont{and} \bibinfo{author}{\bibfnamefont{M.}~\bibnamefont{Sasaki}},
  \bibinfo{journal}{Gen.Rel.Grav.} \textbf{\bibinfo{volume}{44}},
  \bibinfo{pages}{1321} (\bibinfo{year}{2012}), \eprint{1104.2692}.

\bibitem[{\citenamefont{Deser and Woodard}(2013)}]{Deser:2013uya}
\bibinfo{author}{\bibfnamefont{S.}~\bibnamefont{Deser}} \bibnamefont{and}
  \bibinfo{author}{\bibfnamefont{R.}~\bibnamefont{Woodard}},
  \bibinfo{journal}{JCAP} p. \bibinfo{pages}{in press} (\bibinfo{year}{2013}),
  \eprint{1307.6639}.

\bibitem[{\citenamefont{Dodelson and Park}(2013)}]{Dodelson:2013sma}
\bibinfo{author}{\bibfnamefont{S.}~\bibnamefont{Dodelson}} \bibnamefont{and}
  \bibinfo{author}{\bibfnamefont{S.}~\bibnamefont{Park}}
  (\bibinfo{year}{2013}), \eprint{1310.4329}.

\bibitem[{\citenamefont{Woodard}(2014)}]{Woodard:2014iga}
\bibinfo{author}{\bibfnamefont{R.}~\bibnamefont{Woodard}}
  (\bibinfo{year}{2014}), \eprint{1401.0254}.

\bibitem[{\citenamefont{Barvinsky}(2003)}]{Barvinsky:2003kg}
\bibinfo{author}{\bibfnamefont{A.}~\bibnamefont{Barvinsky}},
  \bibinfo{journal}{Phys.Lett.} \textbf{\bibinfo{volume}{B572}},
  \bibinfo{pages}{109} (\bibinfo{year}{2003}), \eprint{hep-th/0304229}.

\bibitem[{\citenamefont{Barvinsky}(2012{\natexlab{a}})}]{Barvinsky:2011hd}
\bibinfo{author}{\bibfnamefont{A.}~\bibnamefont{Barvinsky}},
  \bibinfo{journal}{Phys.Lett.} \textbf{\bibinfo{volume}{B710}},
  \bibinfo{pages}{12} (\bibinfo{year}{2012}{\natexlab{a}}), \eprint{1107.1463}.

\bibitem[{\citenamefont{Barvinsky}(2012{\natexlab{b}})}]{Barvinsky:2011rk}
\bibinfo{author}{\bibfnamefont{A.~O.} \bibnamefont{Barvinsky}},
  \bibinfo{journal}{Phys.Rev.} \textbf{\bibinfo{volume}{D85}},
  \bibinfo{pages}{104018} (\bibinfo{year}{2012}{\natexlab{b}}),
  \eprint{1112.4340}.

\bibitem[{\citenamefont{Koshelev}(2009)}]{Koshelev:2008ie}
\bibinfo{author}{\bibfnamefont{N.}~\bibnamefont{Koshelev}},
  \bibinfo{journal}{Grav.Cosmol.} \textbf{\bibinfo{volume}{15}},
  \bibinfo{pages}{220} (\bibinfo{year}{2009}), \eprint{0809.4927}.

\bibitem[{\citenamefont{Koivisto}(2010)}]{Koivisto:2009jn}
\bibinfo{author}{\bibfnamefont{T.~S.} \bibnamefont{Koivisto}},
  \bibinfo{journal}{AIP Conf.Proc.} \textbf{\bibinfo{volume}{1206}},
  \bibinfo{pages}{79} (\bibinfo{year}{2010}), \eprint{0910.4097}.

\bibitem[{\citenamefont{Caldwell}(2002)}]{Caldwell:1999ew}
\bibinfo{author}{\bibfnamefont{R.}~\bibnamefont{Caldwell}},
  \bibinfo{journal}{Phys.Lett.} \textbf{\bibinfo{volume}{B545}},
  \bibinfo{pages}{23} (\bibinfo{year}{2002}), \eprint{astro-ph/9908168}.

\bibitem[{\citenamefont{Carroll et~al.}(2003)\citenamefont{Carroll, Hoffman,
  and Trodden}}]{Carroll:2003st}
\bibinfo{author}{\bibfnamefont{S.~M.} \bibnamefont{Carroll}},
  \bibinfo{author}{\bibfnamefont{M.}~\bibnamefont{Hoffman}}, \bibnamefont{and}
  \bibinfo{author}{\bibfnamefont{M.}~\bibnamefont{Trodden}},
  \bibinfo{journal}{Phys.Rev.} \textbf{\bibinfo{volume}{D68}},
  \bibinfo{pages}{023509} (\bibinfo{year}{2003}), \eprint{astro-ph/0301273}.

\bibitem[{\citenamefont{Dirian et~al.}(2014)\citenamefont{Dirian, Foffa,
  Khosravi, Kunz, and Maggiore}}]{Dirianetal}
\bibinfo{author}{\bibfnamefont{Y.}~\bibnamefont{Dirian}},
  \bibinfo{author}{\bibfnamefont{S.}~\bibnamefont{Foffa}},
  \bibinfo{author}{\bibfnamefont{N.}~\bibnamefont{Khosravi}},
  \bibinfo{author}{\bibfnamefont{M.}~\bibnamefont{Kunz}}, \bibnamefont{and}
  \bibinfo{author}{\bibfnamefont{M.}~\bibnamefont{Maggiore}},
  \bibinfo{journal}{in preparation}  (\bibinfo{year}{2014}).

\bibitem[{\citenamefont{Jordan}(1986)}]{Jordan:1986ug}
\bibinfo{author}{\bibfnamefont{R.}~\bibnamefont{Jordan}},
  \bibinfo{journal}{Phys.Rev.} \textbf{\bibinfo{volume}{D33}},
  \bibinfo{pages}{444} (\bibinfo{year}{1986}).

\bibitem[{\citenamefont{Calzetta and Hu}(1987)}]{Calzetta:1986ey}
\bibinfo{author}{\bibfnamefont{E.}~\bibnamefont{Calzetta}} \bibnamefont{and}
  \bibinfo{author}{\bibfnamefont{B.}~\bibnamefont{Hu}},
  \bibinfo{journal}{Phys.Rev.} \textbf{\bibinfo{volume}{D35}},
  \bibinfo{pages}{495} (\bibinfo{year}{1987}).

\bibitem[{\citenamefont{Carroll et~al.}(2013)\citenamefont{Carroll,
  Leichenauer, and Pollack}}]{Carroll:2013oxa}
\bibinfo{author}{\bibfnamefont{S.~M.} \bibnamefont{Carroll}},
  \bibinfo{author}{\bibfnamefont{S.}~\bibnamefont{Leichenauer}},
  \bibnamefont{and} \bibinfo{author}{\bibfnamefont{J.}~\bibnamefont{Pollack}}
  (\bibinfo{year}{2013}), \eprint{1310.2920}.

\bibitem[{\citenamefont{Jaccard
  et~al.}(2013{\natexlab{b}})\citenamefont{Jaccard, Maggiore, and
  Mitsou}}]{Jaccard:2012ut}
\bibinfo{author}{\bibfnamefont{M.}~\bibnamefont{Jaccard}},
  \bibinfo{author}{\bibfnamefont{M.}~\bibnamefont{Maggiore}}, \bibnamefont{and}
  \bibinfo{author}{\bibfnamefont{E.}~\bibnamefont{Mitsou}},
  \bibinfo{journal}{Phys.Rev.} \textbf{\bibinfo{volume}{D87}},
  \bibinfo{pages}{044017} (\bibinfo{year}{2013}{\natexlab{b}}),
  \eprint{1211.1562}.

\bibitem[{\citenamefont{Deser et~al.}(1966)\citenamefont{Deser, Trubatch, and
  Trubatch}}]{Deser:1966zzb}
\bibinfo{author}{\bibfnamefont{S.}~\bibnamefont{Deser}},
  \bibinfo{author}{\bibfnamefont{J.}~\bibnamefont{Trubatch}}, \bibnamefont{and}
  \bibinfo{author}{\bibfnamefont{S.}~\bibnamefont{Trubatch}},
  \bibinfo{journal}{Can.J.Phys.} \textbf{\bibinfo{volume}{44}},
  \bibinfo{pages}{1715} (\bibinfo{year}{1966}).

\bibitem[{\citenamefont{Alberte et~al.}(2010)\citenamefont{Alberte,
  Chamseddine, and Mukhanov}}]{Alberte:2010it}
\bibinfo{author}{\bibfnamefont{L.}~\bibnamefont{Alberte}},
  \bibinfo{author}{\bibfnamefont{A.~H.} \bibnamefont{Chamseddine}},
  \bibnamefont{and} \bibinfo{author}{\bibfnamefont{V.}~\bibnamefont{Mukhanov}},
  \bibinfo{journal}{JHEP} \textbf{\bibinfo{volume}{1012}}, \bibinfo{pages}{023}
  (\bibinfo{year}{2010}), \eprint{1008.5132}.

\bibitem[{\citenamefont{Carroll}(1997)}]{Carroll:1997ar}
\bibinfo{author}{\bibfnamefont{S.~M.} \bibnamefont{Carroll}}
  (\bibinfo{year}{1997}), \eprint{gr-qc/9712019}.

\bibitem[{\citenamefont{Vainshtein}(1972)}]{Vainshtein:1972sx}
\bibinfo{author}{\bibfnamefont{A.}~\bibnamefont{Vainshtein}},
  \bibinfo{journal}{Phys.Lett.} \textbf{\bibinfo{volume}{B39}},
  \bibinfo{pages}{393} (\bibinfo{year}{1972}).

\bibitem[{\citenamefont{Chevallier and Polarski}(2001)}]{Chevallier:2000qy}
\bibinfo{author}{\bibfnamefont{M.}~\bibnamefont{Chevallier}} \bibnamefont{and}
  \bibinfo{author}{\bibfnamefont{D.}~\bibnamefont{Polarski}},
  \bibinfo{journal}{Int.J.Mod.Phys.} \textbf{\bibinfo{volume}{D10}},
  \bibinfo{pages}{213} (\bibinfo{year}{2001}), \eprint{gr-qc/0009008}.

\bibitem[{\citenamefont{Linder}(2003)}]{Linder:2002et}
\bibinfo{author}{\bibfnamefont{E.~V.} \bibnamefont{Linder}},
  \bibinfo{journal}{Phys.Rev.Lett.} \textbf{\bibinfo{volume}{90}},
  \bibinfo{pages}{091301} (\bibinfo{year}{2003}), \eprint{astro-ph/0208512}.

\bibitem[{\citenamefont{Ade et~al.}(2013)}]{Ade:2013zuv}
\bibinfo{author}{\bibfnamefont{P.}~\bibnamefont{Ade}} \bibnamefont{et~al.}
  (\bibinfo{collaboration}{Planck Collaboration}) (\bibinfo{year}{2013}),
  \eprint{1303.5076}.

\bibitem[{\citenamefont{Rest et~al.}(2013)\citenamefont{Rest, Scolnic, Foley,
  Huber, Chornock et~al.}}]{Rest:2013bya}
\bibinfo{author}{\bibfnamefont{A.}~\bibnamefont{Rest}},
  \bibinfo{author}{\bibfnamefont{D.}~\bibnamefont{Scolnic}},
  \bibinfo{author}{\bibfnamefont{R.}~\bibnamefont{Foley}},
  \bibinfo{author}{\bibfnamefont{M.}~\bibnamefont{Huber}},
  \bibinfo{author}{\bibfnamefont{R.}~\bibnamefont{Chornock}},
  \bibnamefont{et~al.} (\bibinfo{year}{2013}), \eprint{1310.3828}.

\bibitem[{\citenamefont{Shafer and Huterer}(2013)}]{Shafer:2013pxa}
\bibinfo{author}{\bibfnamefont{D.~L.} \bibnamefont{Shafer}} \bibnamefont{and}
  \bibinfo{author}{\bibfnamefont{D.}~\bibnamefont{Huterer}}
  (\bibinfo{year}{2013}), \eprint{1312.1688}.

\bibitem[{\citenamefont{Riess et~al.}(2011)}]{Riess:2011yx}
\bibinfo{author}{\bibfnamefont{A.~G.} \bibnamefont{Riess}}
  \bibnamefont{et~al.}, \bibinfo{journal}{Astrophys.J.}
  \textbf{\bibinfo{volume}{730}}, \bibinfo{pages}{119} (\bibinfo{year}{2011}),
  \eprint{1103.2976}.

\bibitem[{\citenamefont{Freedman et~al.}(2012)}]{Freedman:2012ny}
\bibinfo{author}{\bibfnamefont{W.~L.} \bibnamefont{Freedman}}
  \bibnamefont{et~al.}, \bibinfo{journal}{Astrophys.J.}
  \textbf{\bibinfo{volume}{758}}, \bibinfo{pages}{24} (\bibinfo{year}{2012}),
  \eprint{1208.3281}.

\bibitem[{\citenamefont{Efstathiou}(2013)}]{Efstathiou:2013via}
\bibinfo{author}{\bibfnamefont{G.}~\bibnamefont{Efstathiou}}
  (\bibinfo{year}{2013}), \eprint{1311.3461}.

\bibitem[{\citenamefont{Verde et~al.}(2013)\citenamefont{Verde, Protopapas, and
  Jimenez}}]{Verde:2013wza}
\bibinfo{author}{\bibfnamefont{L.}~\bibnamefont{Verde}},
  \bibinfo{author}{\bibfnamefont{P.}~\bibnamefont{Protopapas}},
  \bibnamefont{and} \bibinfo{author}{\bibfnamefont{R.}~\bibnamefont{Jimenez}},
  \bibinfo{journal}{Phys.Dark Univ.} \textbf{\bibinfo{volume}{2}},
  \bibinfo{pages}{166} (\bibinfo{year}{2013}), \eprint{1306.6766}.

\end{thebibliography}
\end{document}